\documentclass{acm_proc_article-sp}

\usepackage{url}
\usepackage{tikz}

\usepackage{listings}
\usepackage{graphicx}
\usepackage{caption}
\usepackage{amsmath}
\usepackage{subcaption}
\usetikzlibrary{positioning,fit,calc}
\usetikzlibrary{shapes,arrows}
\usetikzlibrary{shapes.multipart}

\begin{document}

\title{Chaotic Memory Randomization for Securing Embedded Systems}

%
%
%
%
%

\numberofauthors{2} 
%
\author{
%
%
\alignauthor
Peter Henderson\\
       \affaddr{Department of Electrical and Computer Engineering}\\
       \affaddr{McGill University}\\
       \affaddr{Montreal, Quebec}\\
       \email{peter.henderson@mail.mcgill.ca}
\alignauthor
Muthucumaru Maheswaran\\
       \affaddr{School of Computer Science}\\
       \affaddr{McGill University}\\
       \affaddr{Montreal, Quebec}\\
       \email{maheswar@cs.mcgill.ca}
}

\maketitle
\begin{abstract}
Embedded systems permeate through nearly all aspects of modern society. From cars to refrigerators to nuclear refineries, securing these systems has never been more important. Intrusions, such as the Stuxnet malware which broke the centrifuges in Iran's Natanz refinery, can be catastrophic to not only the infected systems, but even to the wellbeing of the surrounding population. Modern day protection mechanisms for these embedded systems generally look only at protecting the network layer, and those that try to discover malware already existing on a system typically aren't efficient enough to run on a standalone embedded system. As such, we present a novel way to ensure that no malware has been inserted into an embedded system. We chaotically randomize the entire memory space of the application, interspersing watchdog-monitor programs throughout, to monitor that the core application hasn't been infiltrated. By validating the original program through conventional methods and creating a clean reset, we can ensure that any inserted malware is purged from the system with minimal effect on the given system. We also present a software prototype to validate the possibility of this approach, but given the limitations and vulnerabilities of the prototype, we also suggest a hardware alternative to the system. 
\end{abstract}

 

\section{Introduction}
Networked embedded and constrained devices (the so called Internet of Things) now pervade almost every aspect of modern technologies. These devices range from household appliances, such as refrigerators, to industrial machinery, such as robotic arms used in factories. The need for protecting these constrained devices has never been more important. A compromised microcontroller can be used for a myriad of malevolent activities including denial of service attacks, fraudulent surveillance, and even the hijacking or destruction of potentially dangerous machinery. In the industrial sector, a clear example of the catastrophic consequences of a compromised microcontroller is the event which occurred in Iran's Natanz nuclear refinery. In 2003, the refinery's centrifuges were targeted and destroyed by the Stuxnet virus \cite{5772960}. The consequences of a similar attack on a car, a functioning power plant, or a military grade drone could be devastating to the lives of thousands and could potentially disrupt the world economy.

Even in the everyday lives of many people, compromised devices can affect them without them even realizing it. An example of the need for securing these IoT devices in household appliances is a botnet that Proofpoint researchers recently discovered \cite{proofpoint} with nearly 100,000 compromised constrained devices, including ``smart appliances'' such as refrigerators and televisions. The network of compromised embedded systems sent out billions of spam emails per day, without the owners of these devices even knowing that their devices had been taken over. Though not physically harmful, these sort of malicious activities can lead to larger electricity and data usage bills, and increases the computing capacity of malicious botnets (which can in turn cause further security breaches and unwanted impacts on people and organizations).

While some steps have been taken to protect these types of devices \cite{proofpoint2, mckelvey1999embedded, cui2011defending}, many of these systems focus solely on protecting the network connection layer. If code or a malicious application is already loaded onto the machines, these protection mechanisms become inadequate. This is particularly relevant in highly secure closed-network facilities (such as the Natanz refinery) where attack vectors may be loaded via a simple USB key or the malware may be disguised in such a way that it doesn't present its functionality immediately, avoiding detection. The Stuxnet virus for example, lay dormant and masked its functionality by feeding previously recorded ``normal'' data to the safeguard applications in place on the centrifuge micro-controllers \cite{5772960}.

Take for example a networked automotive microcontroller. These controllers already have fault tolerant behavior which can reset without affecting performance \cite{937830}. If a malicious program is loaded onto an automotive microcontroller without a driver's knowledge, a potentially lethal situation could occur where an attack vector lays dormant until high speeds are reached and then turns off steering capabilities, or worse. In this situation, knowing that a malicious program has been loaded onto the microcontroller and resetting the device from a clean authenticated image is much more preferable to the alternative. As such, we present a ``kernel'' or, in the future, a dedicated hardware unit, that can be used to deeply randomize and execute a program to prevent this very issue. We also present another addition to this Chaotic Program Randomization which can alert an external source of a possible intrusion.

\section{Toward a Secure Embedded System}
The key issue we intend to address with our system is the detection and removal of such malicious code from an embedded system. This includes attacks ranging from code-injection through buffer overflows to simple loading mechanisms such as a USB key. With current program and memory structures there are few feedback mechanisms capable of detecting whether code has been inserted into a normally functioning application or if another application has been loaded onto the controller or embedded system. The mechanisms that do exist, such as performance analysis-based techniques, rarely are applicable to embedded solutions due to size and performance constrains. While Cui et al. \cite{cui2011defending} try a biologically inspired approach to interleave a ``symbiote'' into existing operating systems, this approach still does not guarantee protection for vulnerabilities or protect against application layer insecurities such as the ``Heartbleed'' vulnerability \cite{heartbleed}.

We propose here the initial steps toward a secure ``Operating Kernel'', which can not only verify that an application is trusted before launching it, but also provide a basic level of memory encryption to protect against memory exploitations at the application layer. We do so through the chaotic randomization of the system's program memory. These same principles can be applied - with modifications - to even secure server or personal computing operating systems. First, we present a software-based prototype of the solution to show the feasibility of chaotically randomizing the program memory space. This solution still contains vulnerabilities that can be exploited and has significant performance setbacks. To remedy these problems, we then suggest a hardware-assisted alternative which would effectively solve the issues of the software only solution and allow for progress toward an efficient and secure embedded system. 

To address the key issue, we make the general assumption that a program loaded onto the embedded system during a boot is a trusted image of the application. This assumption is made knowing that modern trusted computing techniques are capable of secure verification in networked devices. Enck et al. \cite{Enck}, for example, suggest a lightweight verification technique for mobile phones that, with modification, would be applicable to this scenario and the ARM TrustZone technology already does this to some degree \cite{5751382}. The AEGIS IBM architecture also is a similarly envisioned secure bootstrapping mechanism for their operating system which would be sufficient in completing this task \cite{arbaugh1997secure}. Given this assumption, we then attempt to address the issue of malicious code either injected or loaded onto the embedded system through various mechanisms, post-system startup. 

Though Address Space Layout Randomization (ASLR) is a widely used deterrent from buffer overflow attacks in operating systems, it has proven \cite{shacham2004effectiveness} that a determined attacker could easily circumvent the defense offered. Even fine-grained ASLR \cite{kil2006address}, a more refined randomization technique which randomizes not only applications in memory but functions within the applications itself, has resulted in systems that can still be compromised through buffer overflow attacks, albeit with much more complex attack systems \cite{6547134}. According to NIST \cite{NIST}, buffer overflow attacks accounted for 14.64\% of vulnerabilities in 2013 despite the introduction of fine-grained ASLR. Few other vulnerabilities came close to this percentage. As such, protecting against these forms of vulnerabilities remains an important gap that has yet to be solved completely. Additionally, this form of randomization, even done at the data level, still does little to protect from exploitation of memory-based vulnerabilities, such as the Heartbleed vulnerability in OpenSSL encryption\cite{heartbleed}.

Despite best efforts, attack vectors will be found in one form or another in software. It is important to know when a system has already been compromised, and to ensure that such vulnerabilities will not affect the end-user significantly. Since even something as simple as a new application being uploaded from a USB key can be considered an attack vector in closed network facilities - as in the probable initial introduction of the Stuxnet malware \cite{5772960} - we aim to determine a method which could prevent any unintended application from being loaded onto a micro-controller. As aforementioned, current malware detection mechanisms \cite{idika2007survey} are minimally effective in real time - particularly if the malware, as in the Stuxnet virus, hides its activity or remains dormant - and generally aren't applicable to embedded systems, so we stray away from typical detection models.

\section{General Solution Overview}

To ensure that a rogue application cannot be loaded onto a constrained device without discovery, we develop a system inspired by the underlying randomization principle of fine-grained ASLR \cite{kil2006address}. In our scheme, we use a chaotic randomization technique to scramble the program memory at the instruction level, rather than the function or class level. This not only allows for a minimal level of program encryption, but also gives an opportunity to create a system which can detect malicious code insertions. 

The detection of the malicious insertions comes in the form of a watcher-monitor system between the ``kernel'' and several small ``monitor'' programs interleaved with the application in memory. This system bares a loose resemblance to Cui et al.'s symbiotic method for monitoring and protecting against intrusions, and could even be used to carry out swift counter attacks or protection measures other than a reset as is described in \cite{cui2011defending}.


\subsection{Chaotic Memory Randomization}

As aforementioned, the core of our secure system revolves around the Chaotic Memory Randomization (CMR) system. While ASLR provides random pads between applications and even ASLP randomizes functions within a program, our approach randomizes the entire program address space and homogenizes the distribution (see Figure 1) on an assembly instruction level. If multiple programs are loaded in memory, they are interleaved together, so that a conventional Program Counter cannot step through either program and any spy program or memory exploitation will be rendered useless. This interleaving of programs not only becomes useful for encrypting the program memory at a basic level, but allows us to intersperse monitors throughout the other programs to alert us of malicious code insertion (as is discussed later and presented in Figure 2). A chaotic decryption kernel then feeds the proper instruction sequence to the program counter to accurately step through the applications on the desired embedded system.

Our initial method for randomization is based in Arnold's cat map. Though better chaotic randomization methods may be developed, this method suits the purposes of CMR and its simplicity is welcomed for smaller co-processors which could be used in the hardware alternatives we later propose.

\subsection{Arnold's Cat Map}
This particular chaotic encryption method was first used by V. Arnold \cite{arnold1968ergodic} for the purpose of image stenography and encryption. Using simple matrix rules, shown in Eq. (1), Arnold successfully showed that a reversible chaotic encryption was possible such that by iterating $s$ times a completely unrecognizable image was created, but it is possible to return to the original image from a totally scrambled image by iterating another $z$ times through Eq. (1) - or more simply by multiplying the matrix to the $k$ power. Internally within the matrix, different values can be used in for $p$ and $q$ to further enhance and differentiate the randomization process. For the purposes of randomizing instructions, we assume this system of chaotic encryption will be secure enough to avoid backtracking. This assumption takes into account that an attacker likely doesn't know what the original image (or instruction sequence) should be. Even if the attacker did in fact manage to learn the key, inserting an attack vector to properly execute would still be extremely difficult without causing a reset of the embedded device through the watchdog in the kernel.

\subsection{Application of the Cat Map}
For our purposes, we convert each instruction address to a 2-dimensional coordinate, apply the Arnold's algorithm (see Eq. (1)) and convert the address back to a 1-dimensional space. This results in a mapping which can be used by the decryption ``kernel'' previously mentioned. The kernel is passed a key set ($k, p, q$) which is used to apply Arnold's cat map algorithm to find the real memory address of a program instruction based on an internal virtual program counter (VPC). Any jumps within the original application update the internal VPC and calculate the real address in the randomized memory.

\begin{equation}
\begin{pmatrix} x' \\ y' \end{pmatrix} =
                \begin{pmatrix} 1 & p \\ q & pq+1 \end{pmatrix}^{k} \begin{pmatrix} x \\ y \end{pmatrix} \mod N \;.
\end{equation}

\begin{figure}
\begin{subfigure}[b]{0.3\textwidth}

\begin{tikzpicture}[>=latex,font=\sffamily,every node/.style={minimum width=1cm,minimum height=1.5em,outer sep=0pt,draw=black,fill=blue!40,semithick}]
        \node at (0,0) (A) {1};
        \node [anchor=west] at (A.east) (B) {2};
        \node [anchor=west] at (B.east) (C) {3};
        \node [anchor=west] at (C.east) (D) {...};
        \node [anchor=west] at (D.east) (E) {98};
        \node [anchor=west] at (E.east) (F) {99};
        \node [anchor=west] at (F.east) (G) {100};
\end{tikzpicture}
  \caption{Not Randomized}

\end{subfigure}

\begin{subfigure}[b]{0.3\textwidth}

\begin{tikzpicture}[>=latex,font=\sffamily,every node/.style={minimum width=1cm,minimum height=1.5em,outer sep=0pt,draw=black,fill=blue!40,semithick}]
        \node at (0,0) (A) {1};
        \node [anchor=west] at (A.east) (B) {99};
        \node [anchor=west] at (B.east) (C) {73};
        \node [anchor=west] at (C.east) (D) {...};
        \node [anchor=west] at (D.east) (E) {82};
        \node [anchor=west] at (E.east) (F) {6};
        \node [anchor=west] at (F.east) (G) {15};
\end{tikzpicture}
  \caption{Randomized}

\end{subfigure}

\caption{Randomization System}

\end{figure}

\subsection{Monitor System}
Several attempts \cite{reeves2011lightweight, arora2005secure, shafiq2008embedded, cui2011defending} have been made at creating efficient monitoring devices for determining if malware has been loaded onto an embedded system. Most, with the exception of Cui et al., of these systems involve either modeling performance metrics of the embedded system and comparing them against previous run times or theoretical run time bounds. This is imprecise and often could be ineffective in the case where the malware remains dormant or has built in mechanisms for avoiding such detection. We instead focus our attention on ``tripwire'' mechanisms which we refer to as watcher-monitor programs. 

To create this tripwire alert system with our chaotically encrypted memory solution, we interleave monitor programs throughout the randomized application (as Cui et al. interleave their symbiote). As seen in Figure 3a, in an unscrambled version of the assembly instructions, the application is placed in linear sequence before a monitor. In the chaotically randomized memory, the monitor and application are interspersed (Figure 3b). As presented in Figure 2, a watchdog program in the kernel pings various monitors interleaved within application memory. 

\begin{figure}
\begin{tikzpicture}[x=1.25in,y=0.65cm]
  \foreach \x in {0,1,...,10}
  {
    \node[inner sep=1pt] (BL\x) at (0,-\x)   {};
    \node[inner sep=1pt] (UR\x) at (1,-\x+1) {};
    \node (MM\x) at ($(BL\x)!0.5!(UR\x)$) {};
    \draw (BL\x) rectangle (UR\x);
    \edef\memoryNumber{\number\numexpr276314+\x\relax}
    \node[anchor=south east] at (BL\x.north west) {\memoryNumber};
  }

  \node (MEMLOC) at ($(MM0)+(0,2cm)$) { \parbox{2cm}{\centering Memory Location} };
  \node (ADDRESS) at ($(MEMLOC)-(0.75,0)$) {\parbox{2cm}{\centering Address} };
  \draw [-] (MEMLOC) -- (MM0|-UR0);
  \draw [-] (ADDRESS) -- ($(MM0|-UR0)-(0.75,0)$);


  \node at (MM0) {Monitor 13};
  \node at (MM1) {Application 125};
  \node at (MM2) {nop};
  \node at (MM3) {Application 37};
  \node at (MM4) {Monitor 5};
  \node at (MM5) {Application 119};
  \node at (MM6) {Application 93};
  \node at (MM7) {nop};
  \node at (MM8) {Application 3};
  \node at (MM9) {Monitor 22};
  \node at (MM10){Application 135};

  \node[draw,anchor=west]  (1byte) at ($(UR10)+(0.5,0)$) { \parbox{1.5cm}{\centering Watchdog-Kernel}};
  \draw [-] (1byte) -- ($(1byte|-UR5)-(0,0.5)$) -- ($(UR5)-(0,0.5)$);

\end{tikzpicture}
\caption{Kernel pinging watchdog}
\end{figure}
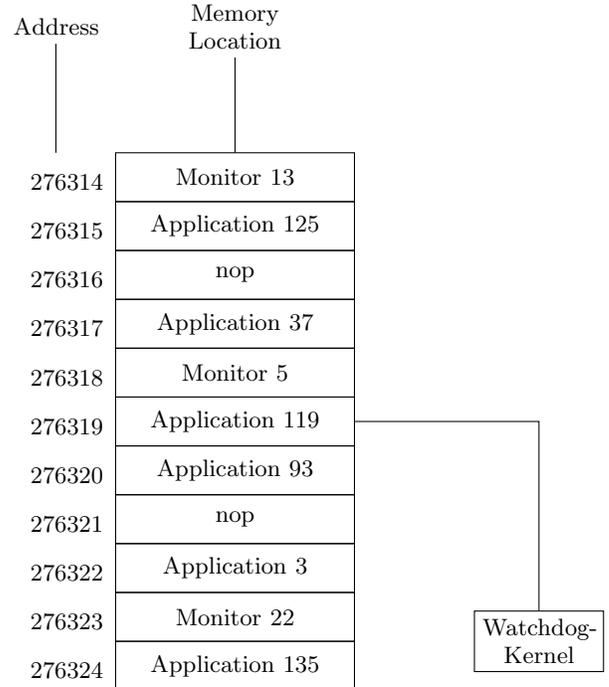

Provided that an acceptable, $m$, number of monitor programs are interleaved with the application, it becomes extremely difficult to insert a malicious program between randomized monitor instructions without overwriting them. With a proper randomization method which homogenizes the memory and spreads the instructions accurately, any insertion or loading of an unauthorized program will overwrite a monitor program instruction (see Figure 3c). When the watchdog then pings a monitor which has been overwritten, the monitor will fail to respond and the watchdog will be alerted of a malicious program, causing a reset, alert, or counter-attack mechanism to launch. A visualization of this can be seen in Figure 2. 

The method in which the watchdog ``pings'' these monitors can be approached in several ways. For simplicity in the initial model, we suggest a simple privileged instruction which periodically resets a validation register or value in the kernel. The validation register would expire after several clock cycles and trigger a ping to one of the monitor programs, which would then reset the register within the kernel. In the software solution, this is emulated by a register which keeps track of this validation bit and is reset every $x$ instruction executions. It then triggers a ping to a monitor location, which resets the register to a valid position. This solution clearly has vulnerabilities as the software-only solution generally does (for example if an attacker learned which register is used for validation and was able to execute a privileged instruction to reset it). We later discuss a hardware alternative to this pinging mechanism to address these vulnerabilities. 

\subsection{Trusted Application Loading}
We assume a trusted application loading mechanism from a clean image to reset after possible intrusions. Using a mixture of methods similar to IBM's AEGIS architecture \cite{arbaugh1997secure} and server-based authentication, we can assure that on the kernel's start, the loaded application is authentic and untampered. IBM's 4758 secure co-processor \cite{dyer2001building} also contains several design principles that could be used to ensure trusted loading mechanisms. In this architecture a factory installed hardware-based authentication mechanism sends a certification request to a server, at which point the code is authenticated and loaded onto the device. Without this authentication, code would not be loaded onto the device memory and would be discarded. Upon successful authentication of a new (or initiating) program, we would randomize the memory, with a fresh chaotic key, before it is placed in RAM using the kernel and keep the needed keys to step through the program.

\begin{figure}
\begin{subfigure}[b]{0.3\textwidth}

\begin{tikzpicture}[>=latex,font=\sffamily,every node/.style={minimum width=1cm,minimum height=1.5em,outer sep=0pt,draw=black,fill=blue!40,semithick}]
        \node at (0,0) (A) {a0};
        \node [anchor=west] at (A.east) (B) {a1};
        \node [anchor=west] at (B.east) (C) {a2};
        \node [anchor=west] at (C.east) (D) {...};
        \node [anchor=west] at (D.east) (E) {m0};
        \node [anchor=west] at (E.east) (F) {m1};
        \node [anchor=west] at (F.east) (G) {m2};
        \node [anchor=west] at (G.east) (H) {m3};
\end{tikzpicture}
  \caption{Unscrambled and Separated Application and Monitor}

\end{subfigure}

\begin{subfigure}[b]{0.3\textwidth}

\begin{tikzpicture}[>=latex,font=\sffamily,every node/.style={minimum width=1cm,minimum height=1.5em,outer sep=0pt,draw=black,fill=blue!40,semithick}]
        \node at (0,0) (A) {a0};
        \node [anchor=west] at (A.east) (B) {m3};
        \node [anchor=west] at (B.east) (C) {a13};
        \node [anchor=west] at (C.east) (D) {...};
        \node [anchor=west] at (D.east) (E) {m5};
        \node [anchor=west] at (E.east) (F) {a6};
        \node [anchor=west] at (F.east) (G) {m12};
        \node [anchor=west] at (G.east) (H) {a35};
\end{tikzpicture}
  \caption{Scrambled and Interleaved}

\end{subfigure}

\begin{subfigure}[b]{0.3\textwidth}

\begin{tikzpicture}[>=latex,font=\sffamily,every node/.style={minimum width=1cm,minimum height=1.5em,outer sep=0pt,draw=black,fill=blue!40,semithick}]
        \node at (0,0) (A) {a0};
        \node [anchor=west] at (A.east) (B) {m3};
        \node [anchor=west] at (B.east) (C) {a13};
        \node [anchor=west] at (C.east) (D) {...};
        \node [anchor=west, fill=red] at (D.east) (E) {v0};
        \node [anchor=west, fill=red] at (E.east) (F) {v1};
        \node [anchor=west, fill=red] at (F.east) (G) {v2};
        \node [anchor=west, fill=red] at (G.east) (H) {v3};
\end{tikzpicture}
  \caption{Attack vector insertion}

\end{subfigure}
\caption{Monitor System}
\end{figure}

\section{Common Attacks and \\Counter Measures}
To illustrate the protection properties of the developed system, we look at some of the common attack vectors against embedded systems and discuss how they're thwarted by our system.

\subsection{Code Injection Attacks}
Code injection attacks often exploit vulnerabilities in the application layer to insert malicious code into the execution path. This is also the main type of attack that our system is designed to thwart. Any insertions into the embedded systems program memory space through buffer overflows, exploitation of double-free vulnerability, integer errors, and the exploitation of format string
vulnerabilities would result in a monitor program or even the application being overwritten, triggering a reset or alert to be issued. All are mechanisms for loading unsafe code onto the embedded system. As we describe earlier, the monitor-watchdog system we propose counters these types of attacks by pinging monitors interleaved in the randomized memory space. If an injection attack is attempted either the application itself or a monitor somewhere in the homogenized memory space will be overwritten, setting off a watchdog alert.

Though much more complex, this same principle can be applied to the data memory space to prevent insertion of data and decrease the chances of data manipulation, but is not discussed in this paper.

\subsection{Side Channel Attacks}
Side channel attacks are attacks which analyze various ``side channels'' - such as power consumption of a chip \cite{kocher1999differential} - in an attempt to decipher an encryption key or acquire some other information about the system. Though our system does not necessarily thwart these types attacks, it is important to address these attack vectors because they may be used to find the encryption pattern in which the memory is scrambled. Given that an attacker does in fact manage to successfully use side channel attacks determine the randomization pattern used in the embedded system, this information does very little to help the attacker. Assuming a secure kernel (as will be discussed in the hardware prototype), the attacker must still overwrite the exact locations of subsequent assembly instructions to ensure no monitors are overwritten and the application does not freeze, alerting a watchdog. This task is difficult with conventional code injection attacks considering the amount of precision needed to complete it, and is enough to thwart most, if not all side-channel based attacks.

\subsection{Memory Vulnerabilities (Heartbleed)}
While not a direct intent of our model, when applied to data memory, chaotic randomization adds a simple layer of encryption to data memory. With our model in place, application layer vulnerabilities (such as the Heartbleed vulnerability) which expose data memory would not be as large of a problem. If chunks of data memory were to be transferred as occurred in Heartbleed, the attacker still would not be able to acquire any useful information due to the scrambled nature of the memory. 

\section{Software Prototype}
\subsection{Introduction}
For an initial solution, we use a prototype software-based kernel on an 8-bit AVR microcontroller (Atmel 328p). The purpose of the prototype was to determine if a program could in fact be run successfully while remaining chaotically randomized at the assembly instruction level. The kernel consists of several parts. An internal virtual program counter (VPC) keeps track of the intended execution procedure (i.e. how a typical Program Counter would function). The three remaining parts of the kernel contain several functions: a one time initialization of all necessary starting addresses, a section which calculates the next ``real'' assembly instruction address in program memory based on the VPC, and a section which updates the VPC based on jumps or branches within the application. The application code was preprocessed and pre-randomized before being merged with the kernel hex file and loaded onto the microcontroller (or emulator). The kernel itself, for the prototype, remained unscrambled and in the 0x0000 region of memory, while the scrambled application was placed in a higher location in memory.

\subsection{Application Preprocessing}
The application assembly instructions had to be preprocessed to ensure proper execution through the kernel. Each instruction was interleaved with a jump back to the proper location in the kernel, and nop instructions were added for normalizing spacing in the program (to ensure the proper calculation of memory addresses after scrambling the program since AVR instructions can be of different lengths). Such a set of instructions (the nop, jump back to the kernel, and the application instruction) was considered to be a single ``instruction'' for the purposes of calculating a memory address in the randomized memory space. This following instructions set would thus be considered as one ``instruction'' for calculating a real address in memory.

\begin{lstlisting}
       movw r22,r18
       jmp 0x01ec
       nop ; padding for calculation
\end{lstlisting}

Sublabels were replaced with full-frame labels (i.e. 1: was converted to .S1) and paired with the immediate following instruction for randomization. As a result and instruction sequence such as:

\begin{lstlisting}
    .L8:
        1: brne 1b
\end{lstlisting}

would be converted to a sequence such as:

\begin{lstlisting}
    .L8:
        brne .L8
        jmp 0x01ec
        nop
\end{lstlisting}

This is because the local labels and jumps (i.e. 1b) would be totally useless because of the instruction level randomization of the program. Since full-frame labels were pair with instructions, in this case a jump to .L8 would serve the same purpose as 1b and would jump to the intended location even in a randomized memory space. However, because we want to keep track of and update the VPC in the kernel, unconditional jumps or branches within the program were replaced with a set of instructions which pushed the intended label address onto the stack, jumped to the part of the kernel which updated the VPC, and then continued to the desired label address. Conditional branches were replaced with a branch sequence which either jumped to the normal part of the kernel ($VPC = VPC + 1$) if the condition was not met, or would branch to slightly further in the sequence which would put the intended label address onto the stack and proceed as with unconditional jumps. A modified example of this sequence can be seen as follows:

\begin{lstlisting}
        brne .L5
\end{lstlisting}

becomes:

\begin{lstlisting}
        brne .X1
        jmp 0x01ec
    .X1:
        ldi r28, lo8(gs(.L5))
        push r28
        ldi r28, hi8(gs(.L5))
        push r28
        jmp 0x02D4
\end{lstlisting}

Here, the intended branching address is .L5, which is replaced with a push onto the stack of the address to update the VPC and jumping to the appropriate part of the kernel. Otherwise, the operation continues on to the normal part of the kernel which executes $VPC = VPC + 1$.

After preprocessing, the application was then randomized (using a simple python script with Arnold's Cat Map), linked and then combined with the kernel through a hex merger. The merged hex file was first loaded into an AVR simulator to validate the execution order of the program. Upon achieving correct execution, the program was loaded onto the Arduino to ensure that the proper execution goal was achieved. The successful running of the prototype solution showed the viability of randomizing the program memory at the instruction level while still achieving the desired execution.

\subsection{Execution and Performance Impact}
As expected, the software prototype has a significant performance impact on the execution of even a simple program. The memory impact is large due to the normalization of assembly instruction sizes and the interspersion of jump instructions to return to the kernel. Additionally, performance suffers due to the calculation of the next instruction's memory location at every increment of the VPC as well as the necessary jump back to the kernel to perform the calculations on the VPC.

\subsection{Security Vulnerabilities}
The software prototype is also vulnerable to attacks on the kernel as it remains linear in execution. An attacker could still modify the kernel or learn of the randomization pattern and/or keys to break the functionality of the system. As previously mentioned, we used the software-only prototype as a basis for the hardware which we propose as follows to ensure maximal security.

\section{Hardware Solution}
With Chaotic Memory Randomization, it is extremely difficult, if not impossible, to create a software-only solution which ensures total protection of the processing kernel and application while still keeping performance impacts to a minimum. As such we propose a hardware alternative based on the software model. The hardware-based solution lays its roots in the foundations set forward by IBM's 4758 Secure Cryptoprocessor \cite{dyer2001building} and in Motorola's very first Memory Management Unit. \cite{zolnowsky1984memory} While the software loaded onto a 4758 Secure Cryptoprocessor has been arguably made vulnerable since its release \cite{bond2001attacks, bond2001api}, its hardware and firmware, remain unbreakable without extreme effort (if at all). As such, we use this co-processor as a base for our proposed architecture to ensure protection at the hardware and firmware level. As a secure system is only as secure as its weakest link, the goal of our hardware based solution, which we will refer to generally as a Chaotic Security Coprocessor (CSC), is to strengthen the security of applications on an embedded system while keeping the CSC itself entirely secure. In our hardware proposed solution, we also suggest the use of a Chaotic Memory Unit for data memory reads, though this will be discussed only briefly.

We have several aims in designing the hardware for our CSC unit:

\begin{itemize}
\item Including an authenticated loading and randomization mechanism
\item Efficiently modifying the program counter to chaotically execute a randomized program.
\item Implementing a secure watchdog-monitor system in a secure hardware solution to prevent tampering
\item Chaotically randomizing and mapping the data memory space through a Chaotic MMU
\item Physically securing the hardware unit itself from tampering, as the 4758 does
\end{itemize}

\subsection{General Architecture}
We propose a unit that can either be integrated into existing CPU architectures or as an isolated unit outside of the processor. To illustrate the proposed capabilities of the unit, we will focus mainly on the latter. The unit is comprised of several parts and generally sits between the CPU and fast-access memory. The parts involved include: a loading unit, a memory randomization unit, a random number generator (which can be forgone in CPU integrated implementations in favor of an existing random number generator located in the CPU), small co-processor for address calculations, several caches for mapping and pre-storage of upcoming addresses, and a watchdog ROM. 

\subsection{Authenticated Loading, Randomization Unit, and Random Number Generator}
As with the IBM's 4758 co-processor, we suggest an integrated code-authentication and loading mechanism. In their solution, Dyer et al., implement a trusted root with a ``securely generated root certificate'' which can later be used with a dedicated authentication unit to verify code before it is loaded into executable memory. We suggest the use of this same mechanism for our hardware implementation with some modification. 

The loading unit, authenticating in the same way as the 4758 co-processor, then generates a random chaotic encryption and decryption key pair using an internal Pseudo-Random Number Generator similar to the randomization system used by FreeBSD (which further randomizes a hardware generated seed using Yarrow). The reason for not using a purely hardware based random number generator is due to the built in backdoors that are known to be exploited by government agencies in Intel and Via Technologies hardware based random number generators, which FreeBSD cited as its reason for no longer relying solely on hardware based generators in 2013 \cite{nsabackdoor, freebsd}. 

Generating a key pair, the loading part of the CSC then proceeds to chaotically randomize the program when loading it into memory. In the same way that the program counter of the system finds the next instruction to execute, the loading mechanism finds the next addresses where to place instructions or data in memory based on the randomization strategy and places this address on the bus for loading the data. As in the software prototype, the CSC would stack small monitor programs on top of the application before the randomization process to ensure that the application was sufficiently interlaced with security ``tripwires''. 

\subsection{Co-Processor and Program Execution}
The core of the chaotic processor is the execution strategy for running a randomized program. This strategy uses the internal co-processor to continuously calculate upcoming instruction addresses and store them in a fast cache according to an internal program counter. In systems where minimal impact to the existing architecture is desired, our CSC would essentially replace the Program Counter. After being delivered a key from the loading mechanism, the CSC would then begin to calculate upcoming memory addresses according to an internal Program Counter (referred to from now on as the VPC for consistency with the Software Prototype). These addresses are stored in a fast cache to be fed to the CPU. From here two approaches could be taken: one which minimizes impact to existing architectures, or one which takes a more consolidated approach. To see a visualization of both approaches please see Figures 4 and 5.

\subsubsection{Minimal Impact}
In the minimal impact approach to existing architectures, the CPU puts the address given by the CSC onto the bus, retrieving the instruction to execute. For jumps or branches, the CPU would place the ``virtual'' address on the bus connecting the CSC which would calculate the next logical address and continue on. In this method, the CSC acts exactly as a program counter would and the Chaotic Memory Unit would interact with it to fetch the real memory address, similar to a conventional MMU.

\subsubsection{Consolidated Approach}
Rather than allowing the CPU direct access to memory, in this approach, the CSC wouldn't even give the CPU any memory addresses, it would simply feed the CPU instructions. Any memory requests would go through the CSC. This would not only allow for fast pre-caching of instructions in the CSC, but would also allow for extra security precautions in the future (perhaps including analysis of suspicious instruction sequences before execution would even occur). In this scenario possible subsequent instruction paths for both unconditional and conditional branches could be pre-cached to lessen the impact of recalculating the upcoming addresses. For data memory access in this scenario, we will refer to the Chaotic Memory Unit section. 

\subsection{Internal Watchdog}
As in the proposed general solution, a hardware based internal watchdog in the CSC would occasionally execute monitor programs in the code to ensure that the application had not been compromised. An unsuccessful execution of the monitor would result in an exception being raised and the CSC shutting down the current context, and relaunching the authenticated loader. More specifically, the monitor could be a privileged instruction which would set a ``security'' register within the CSC. The security register would reset with $x$ cycles and trigger a monitor sampling. An example sampling, which could be altered to increase security, could be a hashing mechanism which is passed a key from the watchdog. Upon hashing, the key hashed value would be verified and reset the validation register in the CSC. If the register were not set back to a valid state, the breach protocol would be triggered. If successful the CSC would determine the location of the next monitor to query and continue regular execution. To ensure that the watchdog cannot be tampered with, we place a small ROM with the watchdog executable code on it inside the secure co-processor.

\subsection{Chaotic Memory Unit (CMU)}
To randomize the data memory space as well, we suggest the use of a similar mechanism to commonplace MMUs, but which instead use mappings provided through the secure co-processor located on-board the CSC. The randomization of data memory would be similar to that of program memory, but would occur at an address space level, rather than at the instruction level. This would also circumvent the need for homogenizing the sizes of assembly instructions for address calculations. As with CSC, we propose two solutions.

\subsubsection{Minimal Impact}
In an effort to preserve existing architectures we propose a simple method in which the CPU places the desired memory addresses on the bus to main memory. The CMU intercepts this, uses the CSC to calculate the real address in main memory and then places this on the outgoing bus. The performance impacts of this are obvious as no pre-caching could be done in this situation.

\subsubsection{Consolidated Approach}
In the consolidated approach, we combined the CSC with the CMU to ensure that data memory addresses can be pre-cached along with program addresses. The CSC would interpret instructions beforehand and calculate the needed memory addresses. Those addresses would be fetched and cached before the CPU would even execute the instruction. For example, in the sequence of instructions as follows, the real address locations of $OCR_1$ and $OCR_2$ would have been pre-calculated and pre-cached.

\begin{lstlisting}
       ldi     temp2, 0
       sts     OCR_1, temp2
       ldi     temp2, 1
       sts     OCR_2, temp2
\end{lstlisting}

In this way, when the CPU executes the sts command, it places $OCR_1$ on the address lines to the consolidated CSC/CMU. The CSC/CMU then places the real address on the address lines to main memory from its cache.

\subsection{Physical securing of the CSC}
To protect the hardware itself, we suggest similar mechanisms as in the 4758 cryptoprocessor architecture. In their hardware solution, ratchet locks protect the sensitive memories (containing secret keys) which zero the coprocessor memory without requiring software intervention on any detected tampering (including ``temperature extremes, voltage variation, and radiation'') \cite{dyer2001building}. Since our co-processor does not require any access from the outside world to the processor itself other than inputting a desired logical memory address and outputting the real memory location (and initial authentication of the code base), we are able to use the same tamper prevention techniques to completely lock the co-processor. Using battery-backed RAM, one can ensure that the RAM is erased if tampering is detected and that any access results in the zeroing of any secrets which can only be initiated on a total reset of the hardware system (as in the 4758).

\begin{figure}
\tikzstyle{block} = [draw, fill=blue!20, rectangle, 
    minimum height=3em, minimum width=6em]
\tikzstyle{mblock} = [draw, fill=blue!20, rectangle, minimum height = .5em, minimum width=.5em]
\tikzstyle{sum} = [draw, fill=blue!20, circle, node distance=1cm]
\tikzstyle{input} = [coordinate]
\tikzstyle{output} = [coordinate]
\tikzstyle{pinstyle} = [pin edge={to-,thin,black}]

\begin{tikzpicture}[auto, node distance=3cm,>=latex', double/.style={draw, rectangle split,rectangle split parts=2},
]
    \node [block] (CPU) {CPU};
    \node [block, double, right of=CPU, above of=CPU] (Decoder) {CSC
     \nodepart{second}
            \tikz[node distance=1.5em]{\node [mblock] (MP) {\tiny MicroProcessor};
                  \node [mblock, below of=MP] (CompReg) {\tiny Computational Registers};
                  \node [mblock, below of=CompReg] (VPC) {\tiny Virtual PC};
                  \node [mblock, below of=VPC] (WatchDog) {\tiny WatchDog ROM};
                  \node [mblock, below of=WatchDog] (PCache) { \tiny Instruction Cache};
                  \node [mblock, below of=PCache] (MTable) {\tiny Memory Lookup Table};
                  \node [mblock, below of=MTable] (KeyReg) { \tiny Key Registers};
                  }
      };
    \node [block, below of=Decoder] (MMU) {Chaotic MMU};
    \node [block, below of=Decoder, right of=MMU, above of=MMU] (DMem) {Data Memory};

    \draw [->] (MMU) -- node {} (Decoder);
    \draw [->] (Decoder) -- node {} (MMU);
    \draw [->] (CPU) -- node {} (MMU);
    \draw [->] (MMU) -- node {} (DMem);
    \draw [->] (Decoder) -- node {} (CPU);
    \draw [->] (CPU) -- node {} (Decoder);


\end{tikzpicture}
\caption{Hardware Overview: CSC and CMU in Minimal Impact Setting}
\end{figure}
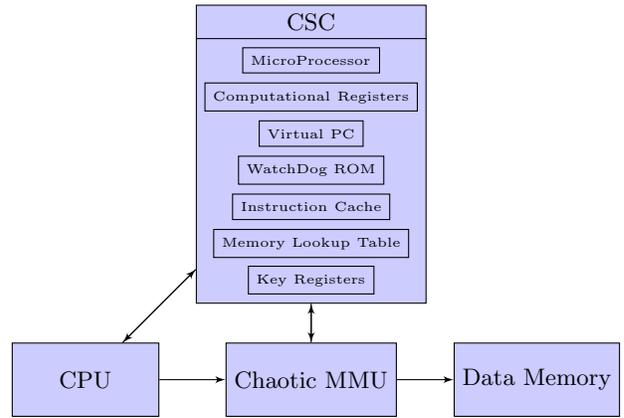

\begin{figure}
\tikzstyle{block} = [draw, fill=blue!20, rectangle, 
    minimum height=3em, minimum width=6em]
\tikzstyle{mblock} = [draw, fill=blue!20, rectangle, minimum height = .5em, minimum width=.5em]
\tikzstyle{sum} = [draw, fill=blue!20, circle, node distance=1cm]
\tikzstyle{input} = [coordinate]
\tikzstyle{output} = [coordinate]
\tikzstyle{pinstyle} = [pin edge={to-,thin,black}]

\begin{tikzpicture}[auto, node distance=3cm,>=latex', double/.style={draw, rectangle split,rectangle split parts=2},
]
    \node [block] (CPU) {CPU};
    \node [block, double, right of=CPU] (Decoder) {CSC
     \nodepart{second}
            \tikz[node distance=1.5em]{\node [mblock] (MP) {\tiny MicroProcessor};
                  \node [mblock, below of=MP] (CompReg) {\tiny Computational Registers};
                  \node [mblock, below of=CompReg] (VPC) {\tiny Virtual PC};
                  \node [mblock, below of=VPC] (WatchDog) {\tiny WatchDog ROM};
                  \node [mblock, below of=WatchDog] (PCache) { \tiny Instruction Cache};
                  \node [mblock, below of=PCache] (MTable) {\tiny Memory Lookup Table};
                  \node [mblock, below of=MTable] (KeyReg) { \tiny Key Registers};
                  \node [mblock, below of=KeyReg] (MMU) {\tiny Chaotic MMU};
                  }
      };
    \node [block, right of=Decoder] (DMem) {Memory};

    \draw [->] (DMem) -- node {} (Decoder);
    \draw [->] (Decoder) -- node {} (DMem);
    \draw [->] (Decoder) -- node {} (CPU);
    \draw [->] (CPU) -- node {} (Decoder);


\end{tikzpicture}
\caption{Hardware Overview: CSC and CMU in Consolidated Setting}
\end{figure}
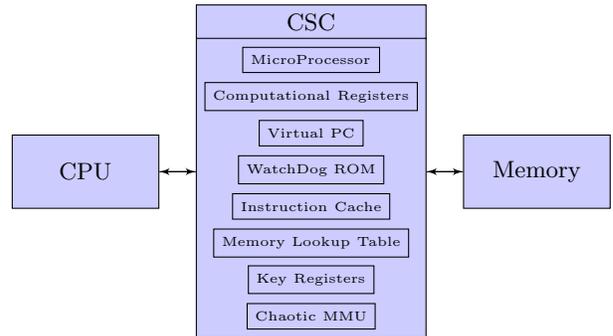

\section{Applications and Conclusions}
While there remains much to be done with the Chaotic Randomization model presented, we set forth an example of what can and needs to be accomplished in rethinking embedded architectures. With the increasing prevalence of embedded systems and the evolving skill and advanced training of malicious attackers, a restructuring of modern day embedded architectures should be sought after to shift mindsets toward securing these systems. While many attacks and errors still occur at the application layer due to the fault of the application programmers, the underlying hardware and memory architectures should do as much as possible to prevent against a catastrophic vulnerability. Chaotic Memory Randomization as a software-only approach cannot and will not totally prevent attacks, yet we demonstrate the possibility of using a totally randomized memory space to successfully execute a program. By keeping security in mind during hardware and chip design processes, steps toward a CSC or something of a similar nature can be taken to ensure that the embedded systems of tomorrow can prevent and detect the insertion of malicious code or alert administrators of malicious activity.

Major applications for such a security co-processor could include secure telecommunications devices that can protect against malware (including keyloggers or wiretaps), autonomous robotics processors which can prevent against ``rogue'' code from being executed, and power plants or other infrastructure facilities among many others. Though our system as described can be applied to many different areas, the focus and where it is most effective is in secure facilities such as the Natanz refinery, where key infrastructure includes many embedded systems that should only run one trusted application. In the case of the Natanz refinery, having the CSC on board the centrifuges could have triggered alarms when the Stuxnet virus was loaded onto the machines and prevented the destruction of refinery equipment. To prevent future such intrusions, particularly in fully functioning nuclear reactors, the development and implementation of a security-centric processing unit, like the CSC, is absolutely necessary.

\bibliographystyle{abbrv}
\bibliography{references}  
%
%
\end{document}